\documentclass[]{kapproc} 
\usepackage{procps} 
\usepackage[dvips]{graphicx}
\usepackage{psfig}
\usepackage{amsmath}
\usepackage{amssymb}
\upperandlowercase
\nochapequationnumber
\let\footnote\savefootnote
\let\footnoterule\savefootnoterule
\let\footnotetext\savefootnotetext

\kluwerbib 
\begin{document}

\articletitle[COSMIC MICROWAVE BACKGROUND]{THE COSMIC
MICROWAVE BACKGROUND\\ ANISOTROPIES: OPEN PROBLEMS}


\author{Enrique Martínez-González}
\affil{Instituto de Física de Cantabria\\
Consejo Superior de Investigaciones Científicas-Universidad de Cantabria}
\email{martinez@ifca.unican.es}
\author{Patricio Vielva}
\affil{Instituto de Física de Cantabria\\
Consejo Superior de Investigaciones Científicas-Universidad de Cantabria}
\email{vielva@ifca.unican.es}


\begin{abstract}
We present a review of some interesting theoretical and observational
aspects of the Cosmic Microwave Background anisotropies.  The standard
inflationary model presents a simple scenario within which the
homogeneity, isotropy and flatness of the universe appear as natural
outcomes and, in addition, fluctuations in the energy density are
originated during the inflationary phase. These seminal density
fluctuations give rise to fluctuations in the temperature of the
Cosmic Microwave Background (CMB) at the decoupling
surface. Afterward, the CMB photons propagate almost freely, with
slight gravitational interactions with the evolving gravitational
field present in the large scale structure (LSS) of the matter
distribution and a low scattering rate with free electrons after the
universe becomes reionized by the first stars and QSOs. These
secondary effects slightly change the shape of the intensity and
polarization angular power spectra of the radiation, the so called
$C_\ell$. The $C_{\ell}$ contain very valuable information on the
parameters characterizing the background model of the universe and
those parametrising the power spectra of both matter density
perturbations and gravitational waves. The extraction of this richness
of information from the $C_\ell$ is complicated by the superposition
of the radiation coming from other Galactic and extragalactic
emissions at microwave frequencies.  In spite of this, in the last few
years data from sensitive experiments have allowed a good
determination of the shape of the $C_\ell$, providing for the first
time a model of the universe very close to spatially flat. In
particular the WMAP first year data, together with other CMB data at
higher resolution and other cosmological data sets, have made
possible to determine the cosmological parameters with a precision of
a few percent.  The most striking aspect of the derived model of the
universe is the unknown nature of most of its energy
contents. This and other open problems in cosmology represent exciting
challenges for the CMB community. The future ESA Planck mission will
undoubtely shed some light on these remaining questions.

\end{abstract}

\section*{Introduction}

In recent years there has been an explosion of cosmological data
allowing a strong progress in the characterization of the cosmological
model of the universe. In particular, recent data of the temperature
of the Cosmic Microwave Background (CMB) have played a crucial role in
the determination of the cosmological parameters. Many experiments
aimed to map the temperature of the CMB have been and are being
carried out, and many others are now being planned to extend present
capabilities in resolution and sensitivity. From the cosmological data
already collected we know that the spatial geometry of the universe is
close to Euclidean, with most of the energy density in the form of the
so called "dark energy (DE)" (with an equation of state close to a
cosmological constant) and most of the matter density in the form of
the so called "cold dark matter (CDM)" (weakly interactive matter with
negligible velocity).  About the primeval density perturbations we
know that they were close to the adiabatic type (constant ratio of the
matter number density to photon density for each matter component) and
Gaussianly distributed with a nearly scale-invariant power spectrum
(i.e. the gravitational potential fluctuations are the same at all
scales).  The generic predictions of inflation ---a very brief episode
of drastic expansion in the very early history of the universe close
to the Planck time--- are consistent with the characteristics of the
model of the universe and of the primeval density fluctuations just
mentioned, the so called "concordance model".  Thus, inflation
provides us with a plausible scenario within which we can understand
the horizon and flatness problems (i.e. why causally disconnected
regions appear statistically similar in the CMB sky and why the
density of the universe is so close to the critical one) and also the
origin of the density perturbations. Although a physical model of
inflation is still lacking, however, it provides us with a
phenomenological scenario within which we can conceptually understand
some fundamental problems related to the origin of the "special"
properties of the universe ---i.e. its homogeneity and critical density---
and its large scale structure (LSS) matter distribution which
otherwise will have to rely on ad hoc initial conditions.  In the past
decade two of the experiments on-board of the NASA COBE satellite,
FIRAS and DMR, established unambiguously the black-body
electromagnetic spectrum of this radiation (Mather et al. 1994, 1999)
and the level and approximate scale-invariant shape of the spectrum of
density fluctuations (Smoot et al. 1992). The former data rebated some
previous results which indicated possible distortions from the
black-body spectrum, establishing the thermal origin of the CMB with a
high precision. The latter confirmed the gravitational instability
theory for the formation of the LSS and determined an initial spectrum
of fluctuations characterized by density fluctuations with
approximately equal amplitude when entering the horizon. These
fundamental results set the basis for the later developments in our
understanding of the universe and opened the era of precision
cosmology.  At the end of the decade several experiments determined
that the universe is close to spatially flat (BOOMERANG, De Bernardis
et al. 2000; MAXIMA, Hanany et al. 2000). More recently the NASA WMAP
satellite (Bennett et al. 2003) nicely confirmed this result and,
together with other higher resolution CMB experiments as well as the
galaxy survey 2dFGRS (Percival et al. 2001), determined the
cosmological parameters with a few percent errors. The combination of
different cosmological data sets not only helps to improve the
precision of the cosmological parameters but, what is more important,
shows the compatibility of the different data sets in the context of
the concordance model. The most important features of this model is
the distribution of the energy content of the universe with about 70\%
of dark energy, 25\% of cold dark matter (CDM) and only 5\% of
baryons.  The detection of the polarization fluctuations by DASI
(Kovac et al. 2002) and the later determination of the
temperature-polarization angular cross-power spectrum by WMAP (Kogut
et al. 2003) has strongly confirmed the concordance model.  Very
recently, there have been several works finding evidences of positive
cross-correlations between the CMB temperature map and the LSS
distribution of galaxies (see e.g. Boughn and Crittenden 2004). These
results represent an independent piece of evidence of the existence
and dominance of dark energy in the recent history of the evolution of
the universe (assuming that the universe is close to flat). Although
the concordance $\Lambda$CDM model represents a good fit to the CMB
data as well as to other cosmological data sets ---namely LSS galaxy
surveys, primordial Big-Bang nucleosynthesis, measurements of the
Hubble constant, SN Ia magnitude-redshift diagram--- there are however
some problems related to the observations and also to their
interpretation.

\begin{figure}[ht]
\includegraphics[angle=90,width=12cm]{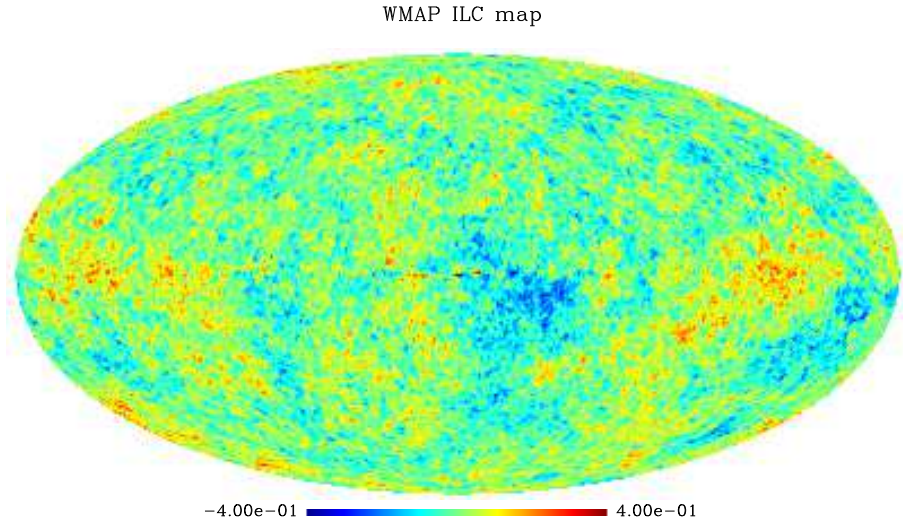}
\caption{\label{fig:WMAP}The CMB as seen by WMAP (data obtained from
the LAMBDA wave page of NASA).}
\end{figure}

The aim of this paper is to review some of the most relevant
theoretical and observational results on the CMB field and to point
out some open problems associated to them. For more detailed
discussions the reader is refereed to the many good reviews written in
the literature, some of them more oriented to the theory (e.g. Hu and
Dodelson 2002, Challinor 2004, 2005, Cabella and Kamionkowski 2004)
and others to the experiments (e.g. Barreiro 2000, Bersanelli, Maino
and Mennella 2002, Mennella et al. 2004).\\ The rest of this review is
as follows. In Section \ref{sec:temperature} we describe the main
properties of the CMB temperature anisotropies and the physical
effects that originate them.  In Section \ref{sec:polarization} the
most relevant properties of the CMB polarization anisotropies are
considered. The cosmological parameters are defined in Section
\ref{sec:cosmological_parameters} and their values determined from
recent CMB data and in combination with other cosmological data sets
are presented in Section \ref{sec:constraints}.  Finally, open
problems in the CMB field are discussed in Section
\ref{sec:open_problems}.

\section{CMB temperature anisotropies}
\label{sec:temperature}
\subsection{Description} 
 
The anisotropies of the CMB ---the fluctuations in the intensity of this
radiation as a function of the direction in the sky--- are interpreted
as a realization of a random field on the sphere. Behind this
interpretation is the idea that we aim to understand and possibly
explain the anisotropies in a statistical manner. Because of the
blackbody spectrum of the CMB, the anisotropies in the intensity are
generally given as temperature ones. Since the temperature
fluctuations are a function of the spherical coordinates it is
convenient to expand them in spherical harmonics,
\begin{equation}
\frac {\Delta T}{T} (\vec n)= \sum_{\ell,m} a_{\ell m} Y_{\ell m}(\vec n),
\end{equation}
where $a_{\ell m}$ are the spherical harmonic coefficients. A very
important quantity is the CMB anisotropy angular power spectrum,
$C_\ell$, the second order moment of the $a_{\ell m}$ defined as
\begin{equation}
<a_{\ell m}a_{\ell' m'}^*> = C_\ell \delta_{\ell\ell'}\delta_{mm'}.
\end{equation}
The null correlation of the harmonic coefficients for different $\ell$
or $m$ is due to the homogeneity and isotropy of the universe
---i.e. the Cosmological Principle assumed as a fundamental pillar in
cosmology and consistent with all of our observations up to date. Here
it is important to notice, however, that the previous equation does
not hold if the universe possesses a nontrivial topology. An important
property of the $C_\ell$ is that if the anisotropies are Gaussian ---as
predicted by Inflation--- then all the statistical information is
contained in it (reported deviations from Gaussianity will be
discussed in section \ref{sec:open_problems}). The correlation
function of the temperature fluctuations, $C(\theta)\equiv <\Delta
T/T(\vec n_1)\Delta T/T(\vec n_2)>$ with $\vec n_1\vec n_2=\cos
(\theta)$, is related to the $C_\ell$ through the Legendre transform
\begin{equation}
C(\theta) = \sum_\ell \frac {2\ell +1}{4\pi} C_\ell 
P_\ell (\cos\theta).
\end{equation}
The isotropy of the field is now reflected in the independence of the
correlation from the direction. Although the two quantities
$C(\theta)$ and $C_\ell$ contain the same information, however the
null correlation of the $C_{\ell's}$ for different values of $\ell$ makes
the latter quantity preferable for cosmological studies. The quantity
which is usually displayed is $\ell (\ell+1) C_\ell /2\pi$, i.e. the
power per logarithmic interval in $\ell$ for large $\ell$. Since the
$\ell =1$ moment is dominated by our motion only moments with $\ell\ge
2$ are considered.

There is a fundamental limitation in the accuracy with which the
angular power spectrum $C_\ell$ can be determined due to the fact that
we can only observe one last-scattering surface ---the error associated
to it is called ``cosmic variance''. Since there are $2\ell+1$
$a_{\ell m}$ coefficients for a given $\ell$ then for Gaussian
temperature fluctuations the cosmic variance is easily calculated
(from the dispersion of a chi-squared distribution with $2\ell+1$
degrees of freedom)
\begin{equation}
\label{eq:corre_error}
\Delta C_\ell = \frac {1} {\sqrt{\ell+0.5}} C_\ell.
\end{equation} 
There are other sources of error which should be added to the cosmic
variance. One is the fraction of the sky $f_{sky}$ covered by an
experiment, which increases the error by a factor $f_{sky}^{-1/2}$
(Scott et al. 1994). Another one is the sensitivity of the experiment
whose noise power spectrum adds to the $C_\ell$ of the cosmic signal
in equation  \ref{eq:corre_error}. Finally, there is a source of error
coming from the process to separate the cosmic signal from the other
foregrounds, namely Galactic emissions\footnote{See e.g. Watson et
al. (2005) for evidence of possible anomalous Galactic emission}
(synchrotron, free-free and thermal dust), extragalactic sources,
galaxy clusters and the lensing effect from the LSS. This is usually a
complex task to perform which requires multifrequency observations and
whose error is difficult to estimate a priori. Estimate of the errors
assuming different situations and methodology can be found in, e.g.,
Hobson et al (1998), Tegmark et al. (2000), Bouchet and Gispert
(1999), Vielva et al. (2001), Maino et al. (2002), Herranz et
al. (2002), Martínez-González et al. (2003), Delabrouille et
al. (2003).

In figure \ref{fig:WMAP} the map of temperature anisotropies as
measured by WMAP is shown (Bennett et al. 2003). Although a large
effort has been made to eliminate all foreground contributions
however some residuals from Galactic foregrounds can still be seen in
the central horizontal band of this map. In order to avoid the
introduction of this contamination in the data analysis a mask
covering that and other regions of the map is usually applied. 

\subsection{Physics}

The CMB anisotropies are usually divided in primary and secondary,
depending if they are produced before or after the last-scattering
surface. The primary anisotropies are the most interesting ones to
study the cosmological parameters characterising the universe as well
as its basic matter and energy constituents. The secondary ones are
produced by scattering of the CMB photons from ionized matter ---either
generated after the reionization of the universe or present as hot gas
in the central regions of galaxy clusters and whose interaction with
the CMB photons is known as the Sunyaev-Zeldovich effect (SZ, Sunyaev and
Zeldovich 1972)---, and by gravitational interactions acting on them
during their travel from the last-scattering surface to the observer
---causing both redshift and lensing effects. They are expected to be
sub-dominant, at least in relation to the $C_\ell$, up to $\ell\approx
2500$.

\begin{figure*}[ht]
\begin{center}
\includegraphics[width=8cm]{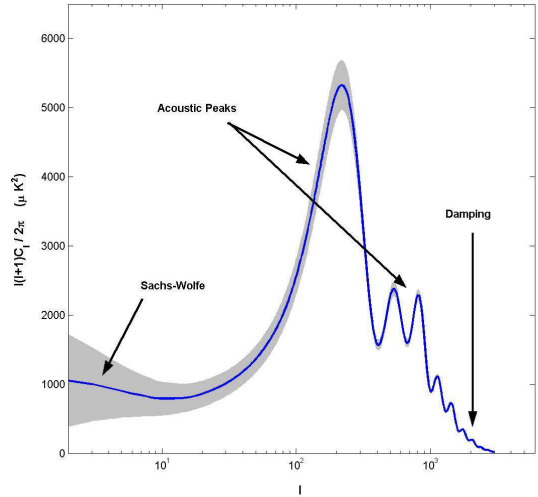}
\caption{\label{fig:spectrum} $C_{\ell}$ for the best fit model given in
Bennett et al. (2003) (see also table \ref{table:WMAP_LSS}). The range
of $\ell$ where the different physical effects dominate is indicated. The 
gray band envolving the $C_{\ell}$ is the unavoidable error due to
the cosmic variance (see text). The
spectrum has been computed using the CMBFAST code (Seljak
and Zaldarriaga 1996).}
\end{center}
\end{figure*}

The main physical effects producing the primary anisotropies and observed in 
the synchronous-comoving gauge (the observer's gauge) can be 
summarized in the following equation (Martínez-González, Sanz
and Silk 1990, Sanz 1997):
\begin{equation}  
{\frac {\Delta T} {T}} (\vec n ) \approx \frac {\phi_{e}(\vec n)}{3} +
\int_e^o \frac {\partial \phi}{\partial t} dt + {\vec n}
\cdot (\vec v_o -\vec v_{e}) + \Big( \frac {\Delta T}{T} (\vec
n)\Big)_{e}  
\end{equation}
The first two terms in the r.h.s. account for the gravitational
redshift suffered by the CMB photons in their travel toward us along
the direction $\vec n$. They are called the Sachs-Wolfe (SW) and
Integrated Sachs-Wolfe (ISW) effects, respectively (Sachs and Wolfe
1967). The third term is a Doppler effect due to the motion of the
emitters at the last-scattering surface. Finally the fourth term is
the temperature fluctuation at that surface.\\ Since the gravitational
field has a large scale of interaction the SW and ISW effects dominate
the angular power spectrum at the largest angular scales ($\ell$
smaller than a few tens) producing an approximate plateau (see
figure  \ref{fig:spectrum}). Notice also that these scales are the most
affected by the cosmic variance (see equation \ref{eq:corre_error}). The
ISW effect has two contributions: an early one before recombination
produced by the imperfect coupling of photons and baryons causing
variations in the gravitational potential with time, and a late one
after recombination due to changes in the gravitational potential with
time. The late ISW effect is produced by the linear
evolution of the large-scale matter distribution at later times if
the universe is different form Einstein-de Sitter. As we will see at
the end of Section \ref{sec:constraints} this is an independent test
from the standard one (involving usually CMB, SN Ia or LSS data sets)
about the existence of a dark energy component dominating the dynamics
of the universe at recent cosmic times.\\ The Doppler effect plus the
temperature fluctuations at the last-scattering surface dominate the
shape of the angular power spectrum at $\ell$ larger than a few
tens. At multipoles above a hundred, the angular power spectrum
exhibits a sequence of oscillations called acoustic peaks. They are
driven by the balance between the gravitational force pulling to
compress over-dense regions and radiation pressure pushing in the
opposite direction. The position and amplitude of these oscillations
are very much determined by the total energy content of the universe
(or equivalently its geometry) and the nature and amount of the
different components.\\ Finally, at large $\ell$ ($\gtrsim 1000$) the
$C_\ell$ power starts to decrease due to the width of the
last-scattering surface and the imperfections of the coupling of the
photon-baryon fluid (Silk effect, Silk 1968).\\ In Figure
\ref{fig:spectrum} one can see the range of $\ell$ where the different
effects dominate the angular power spectrum.

\section{Polarization}
\label{sec:polarization}

Thomson scattering of the radiation generates linear polarization at
the end of recombination when the growth of the mean free path of the
photons allow anisotropies to grow (for a detailed description of
the physics of polarization see, e.g., Challinor 2005). The expected level of
polarization is only of $\approx 5\%$. From the Stokes parameters
$Q,U$ two rotationally invariant quantities can be constructed $E,B$
(often referred to as the E-mode and B-mode, see Zaldarriaga and
Seljak 1997, Kamionkowski et al. 1997). Under parity transformations
$E$ remains unchanged and $B$ changes sign, and therefore the 
cross-power $<a_{\ell m}^B a_{\ell m}^{T*}>=<a_{\ell m}^E
a_{\ell m}^{B*}>=0$. It is for these parity properties that only three
angular power spectra are required to characterise CMB polarization:
$C_\ell^E,C_\ell^B,C_\ell^{TE}$.

\begin{figure*}
\includegraphics[angle=270, width=12cm]{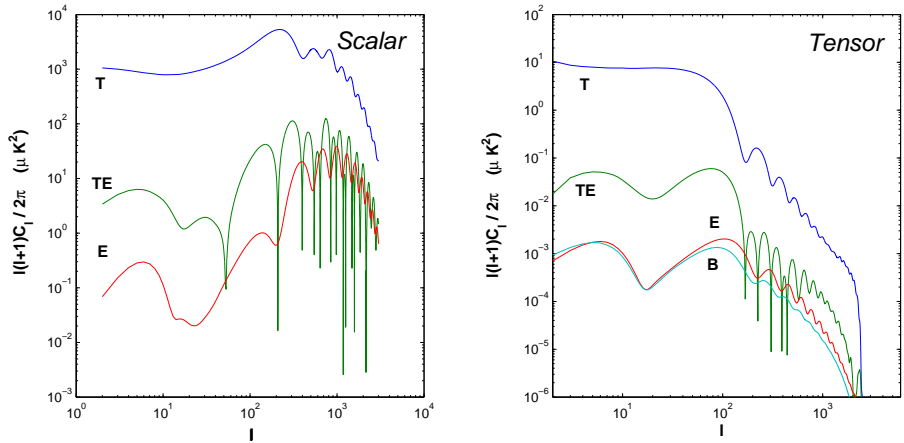}
\caption{\label{fig:spectra_model} Temperature and polarization
angular power spectra for scalar and tensor perturbations. The ratio
tensor/scalar has been chosen to be $r=0.01$. Apart from this
parameter, the vlues for the rest of the parameters have been fixed to
the values of the best fit model given in Bennett et al. 2003 (see
also table \ref{table:WMAP_LSS}). All the spectra have been computed
using the CMBFAST code.}
\end{figure*}

An important property of the E,B decomposition is that whereas the
E-mode polarization can be generated by both density perturbations and
gravitational waves the B-mode can only be generated by the second
ones. Therefore the detection of B-mode polarization is a unique proof
of the existence of primordial gravitational waves, opening the door
to new areas of physics. In particular, it can also be used to impose
strong constraints on the energy scale of inflation and on the shape
of the inflaton potential. A complication comes from the lensing
effect due to the gravitational potential of the LSS which converts
part of the E-mode polarization in B-mode. Fortunately, this effect
dominates over the primary one only at $\ell \gtrsim 100$ leaving the rest
of the spectrum unaltered. All the temperature and polarization
angular power spectra for scalar and tensor perturbations are plotted
in figure \ref{fig:spectra_model}.

As for the temperature angular power spectrum, the accuracy in the
determination of the polarization angular power spectra has also a
fundamental limit imposed by the cosmic variance. For Gaussian
distributed anisotropies the errors in $C_\ell^E,C_\ell^B,C_\ell^{TE}$
are given by:
\begin{equation}
\Delta C_\ell^{E,B} = \frac {1} {\sqrt{\ell+0.5}} C_\ell^{E,B},\ \ \
\Delta C_\ell^{TE} = \frac {1} {\sqrt{\ell+0.5}} \Big( C_\ell^T C_\ell^E +
{C_\ell^{TE}}^2 \Big)^{1/2}.
\end{equation} 
    
There are a number of Galactic and extragalactic foregrounds which
complicate the observation of the CMB polarization. Although their relevance
depends very much on the frequency they are expected to be very harmful,
specially for the B-mode due to its relatively small amplitude. For a
recent estimate of the effect of foregrounds on the polarization
observations see Tucci et al. (2005).   

Only recently the DASI experiment has been able, for the first time,
to detect anisotropies in polarization (Kovac et al. 2002). Afterward, the
WMAP experiment measured the TE angular cross-power spectrum with more
precision and covering a much wider range of scales (Kogut et
al. 2003). More recently the CBI experiment has measured the E angular
power spectrum with more resolution than DASI allowing the detection
of the second, third and fourth acoustic peaks (Redhead et al. 2004b).

In addition to the temperature anisotropies, the anisotropies in
polarization contain very relevant and independent information. In
particular, in the standard model the maxima in the E spectrum are out
of phase with those in the T spectrum due to the fact that polarized
radiation is sensitive to the velocity of the fluid, and the velocity
and density are out of phase in an acoustic wave. This shift is
precisely what has been recently reported by Readhead et al. (2004b)
based on CBI data (see figure  \ref{fig:polarization} below). As we will
see later, anisotropies in the polarization are very relevant to
confirm the best fit model given by the temperature data and to
constrain specific parameters as the optical depth to which they are
very sensitive.

\section{Cosmological parameters}
\label{sec:cosmological_parameters}

There are a number of cosmological parameters that account for very
different fundamental physical properties of the universe and that
influence the radiation angular power spectrum in many ways.  These
parameters characterize the background model of the universe (assumed
to be the homogeneous and isotropic Friedmann-Robertson-Walker model),
the primordial scalar and tensor fluctuations and the reionisation
history.  At present around 12 parameters are considered for the data
analysis. 

The parameters characterizing the background model of the universe
are the following:
 
\begin{itemize}

\item {\sl Physical baryonic density, $w_b$}: $w_b = \Omega_b h^2$.

\item {\sl Physical  matter density, $w_m$}: $w_m = \Omega_m h^2$. 
$\Omega_m$ is given by the sum of the baryonic density $\Omega_b$, 
the CDM density $\Omega_{CDM}$ and the neutrino density $\Omega_\nu$.

\item {\sl Physical neutrino density, $w_\nu$}: $w_\nu = \Omega_\nu h^2$ 
(up to now only upper limits are found for this parameter).

\item {\sl Dark energy equation of state parameter, $w$}:  
$w\equiv p_{DE}/\rho_{DE}$.

\item {\sl Dark energy density, $\Omega_{DE}$}: In case $w$ were
constant and took the value $-1$ then the dark energy takes the form
of a cosmological constant and its energy contribution is represented
by $\Omega_{\Lambda}$.

\item {\sl Hubble constant, $h$}: $h\equiv H_0/100$km
s$^{-1}$ Mpc$^{-1}$. 

\end{itemize}

The reionisation history of the universe influences the CMB by a
single parameter:

\begin{itemize}

\item{\sl Optical depth, $\tau$}: $\tau = \sigma_T \int^{t_0}_{t_r} n_e(t) 
dt$, where $\sigma_T$ is the Thompson cross-section and $n_e(t)$ is the 
electron number density as a function of time.
 
\end{itemize}

The cosmological parameters that characterize the matter and
gravitational waves primordial power spectra are:

\begin{itemize}

\item {\sl Amplitude of the primordial scalar power spectrum,
$A_s$}:\\ $P_s(k) =A_s (k/k_0)^{n_s}$, where $k_0=0.05$Mpc$^{-1}$.  

\item {\sl Scalar spectral index, $n_s$}.

\item {\sl Running index, $\alpha$}: $\alpha=dn_s/d\ln k$. It accounts for the
deviations from a pure power-law. Its value is normally determined at the scale
$k_0=0.05$Mpc$^{-1}$.

\item {\sl Tensor-to-scalar ratio, $r$}: $r=A_t/A_s$. 

\item {\sl Tensor spectral index, $n_t$}: from the consistency relation of 
inflation it is normally assumed that $n_t=-r/8$.

\end{itemize}

Besides those parameters there are two possible types of primordial
matter density fluctuations: adiabatic (the entropy per particle is
preserved) and isocurvature (matter fluctuations compensate those of
the radiation conserving total energy density). As commented in the
Introduction the standard model of inflation predicts fluctuations of
the adiabatic type.\\ 
Analyses that combine CMB with LSS require an
additional parameter accounting for the bias $b$ of the galaxy density
respect to the matter density.

\begin{figure*}
\includegraphics[angle=270,width=12cm]{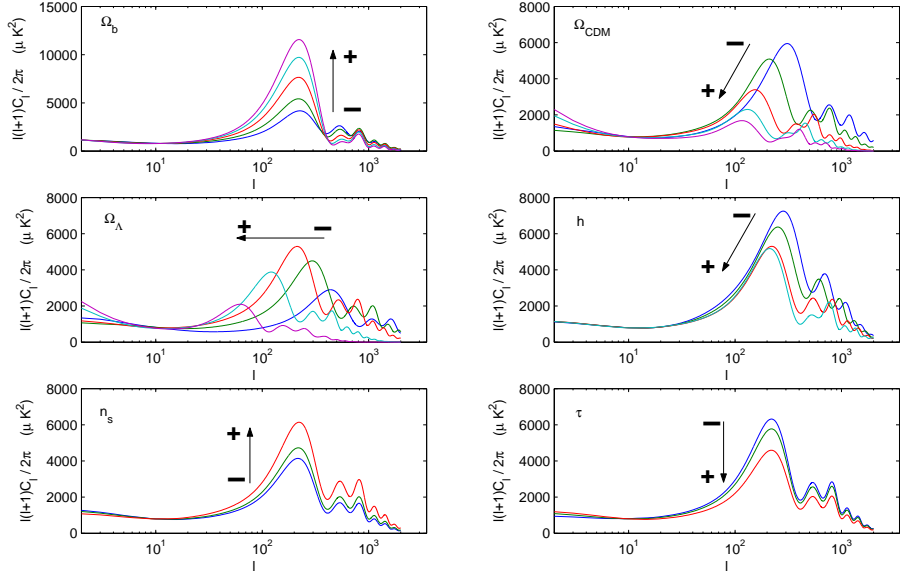}
\caption{\label{fig:dependence} Dependence of $C_{\ell}$ on some
relevant cosmological parameters (The $C_{\ell}$ has been produced
with the CMBFAST code.}
\end{figure*}

\begin{figure*}[h]
\includegraphics[angle=270,width=12cm]{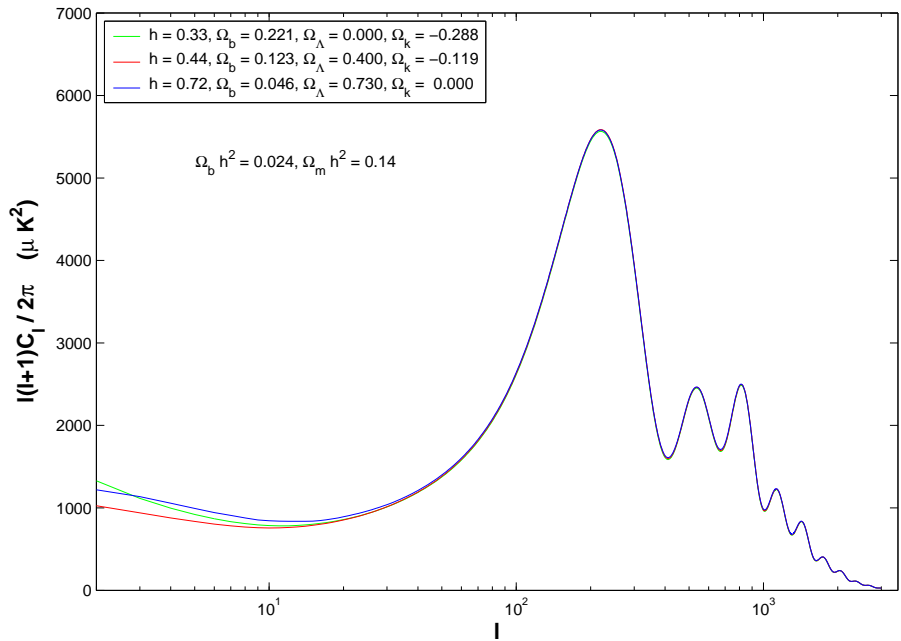}
\caption{\label{fig:degeneration} Geometric degeneracy for three
values of the curvature $\Omega_k$. Apart from the parameters
given in the figure the rest of the parameters have been fixed to the
best fit model given in Bennett et al. (2003). The three
spectra have been computed using the CMBFAST code.}
\end{figure*}

In figure \ref{fig:dependence} we show the changes in $C_\ell$
produced by the variation of some of the most relevant cosmological
parameters. The many different ways in which the $C_\ell$ can vary
with the parameters produces degeneracies complicating their accurate
determination. In particular, there is a well known geometric
degeneracy involving the matter and dark energy densities. This is
shown in figure \ref{fig:degeneration} where almost identical angular
power spectra are obtained for three different values of the curvature
($\Omega_k=1-\Omega_m-\Omega_{DE}$). This example illustrates the need
of including in the analysis additional cosmological data sets (like
SN Ia, LSS, cluster density, CMB polarization, ...) to break the
degeneracies.

\section{Cosmological constraints}
\label{sec:constraints}

In recent years there has been an explosion of cosmological data which
have made possible a strong advance in the determination of the
cosmological parameters. Below we summarize the main results for CMB
data alone and when combined with other cosmological data sets.

\subsection{WMAP and higher resolution CMB experiments}

The WMAP data alone is able to put strong constraints on some
cosmological parameters when some priors are assumed in the analysis
(Spergel et al. 2003).  Table \ref{table:WMAP} summarizes the results
when a flat universe is assumed, the prior $\tau < 0.3$ is imposed on
the optical depth and only 6 parameters are considered. It is
interesting to notice that the Einstein-de Sitter universe (i.e. a spatially
flat universe with null dark energy) is rejected at a very high
confidence level. Besides, the value of the optical depth parameter
$\tau$ is essentially determined by the TE angular cross-power
spectrum.

The WMAP data also test the type of the primordial fluctuations. The
clear detection of the first acoustic peaks as well as the detection
of the TE cross-correlation imply that the fluctuations were primarily  
adiabatic, in agreement with the standard inflationary model.

\begin{table*}[ht]
\begin{center}
\caption[]{\label{table:WMAP} Cosmological parameters using only WMAP
data. In the fit the universe is assumed to be spatially flat and the
value of the optical depth is constrained to $\tau<0.3$ (from Spergel
et al. 2003)}
\begin{tabular}{cc}
\sphline
\it Parameter& \it Values (68\% CL)\cr
\sphline
$w_b$ & 0.024$\pm$0.001\cr
$w_m$ & 0.14$\pm$0.02\cr
$h$ & 0.72$\pm$0.05\cr
$A_s$ & 0.9$\pm$0.1\cr
$\tau$ & 0.166$_{-0.071}^{+0.076}$\cr
$n_s$ & 0.99$\pm$0.04\cr
\sphline
\end{tabular}
\end{center}
\end{table*}

\begin{figure*}
\includegraphics[angle=270,width=12cm]{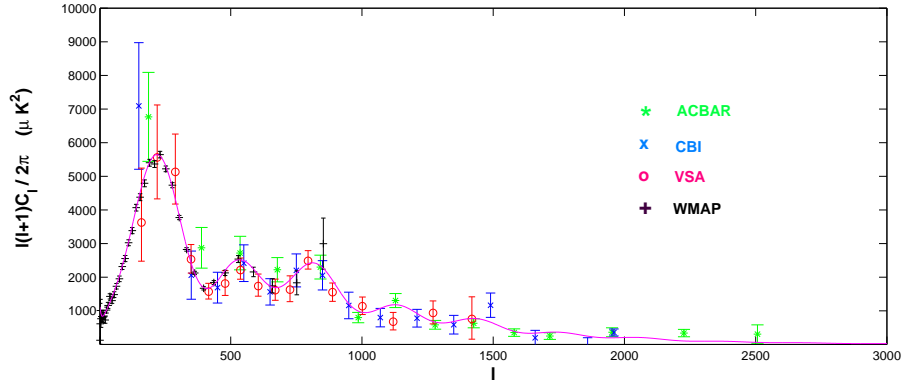}
\caption{\label{fig:temperature} $C_{\ell}$ measured by WMAP, ACBAR,
CBI and VSA, together with the best fit model given by Bennett et
al. (2003). The values of the parameters are listed in table
\ref{table:WMAP_LSS}.}
\end{figure*}

\begin{figure*}[h]
\includegraphics[angle=270,width=12cm]{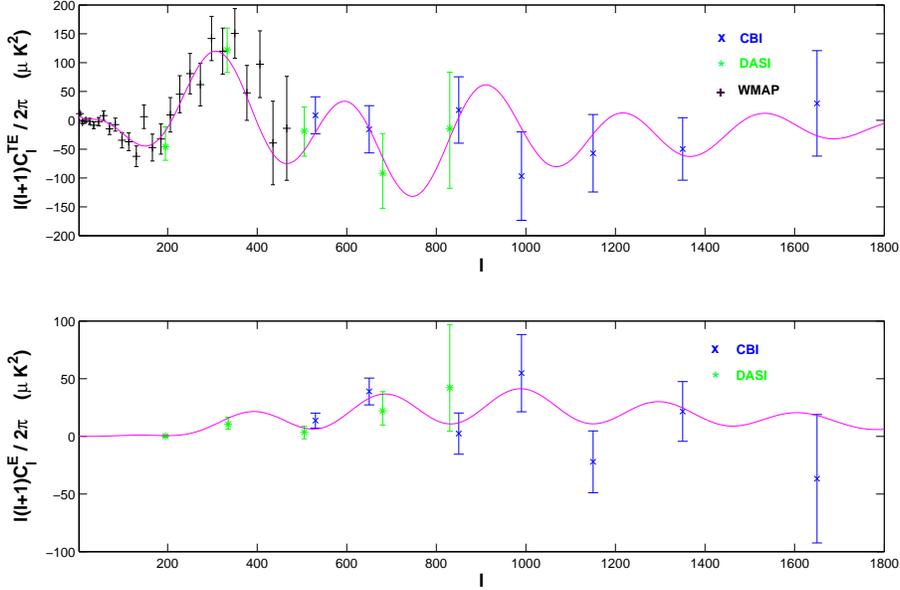}
\caption{\label{fig:polarization} TE and E angular
power spectra measured by DASI, WMAP and CBI (TE) and DASI and CBI
(E). Also plotted is the best fit model given by Bennett et
al. (2003). The values of the parameters are listed in
\ref{table:WMAP_LSS}.}
\end{figure*}

The situation is improved if temperature data from high resolution CMB
experiments is included in the analysis (Spergel et al. 2003,
Dickinson et al. 2004). In figure \ref{fig:temperature} the $C_\ell$
obtained from the experiments WMAP (Hinshaw et al. 2003), ACBAR (Kuo
et al. 2004), CBI (Readhead et al. 2004a) and VSA (Dickinson et
al. 2004) is shown. Although the polarization measurements are not yet
sensitive enough to significantly improve the constraints already
derived from the temperature one (except for $\tau$), however the
measured peaks in the E-mode $C_{\ell}$, which are out of phase with
the temperature $C_{\ell}$ ones, suppose an independent evidence of
the standard model and, more specifically of the adiabatic type of the
primordial matter density fluctuations.  The TE angular cross-power
spectrum from DASI (Leitch et al. 2004), WMAP (Kogut et al. 2003) and
CBI (Readhead et al.  2004b) as well as the E-mode polarization
angular power spectrum from DASI and CBI are shown in
figure  \ref{fig:polarization}.

\subsection{Combination with other cosmological data sets}

\begin{figure*}[h]
\begin{center}
\includegraphics[width=8cm]{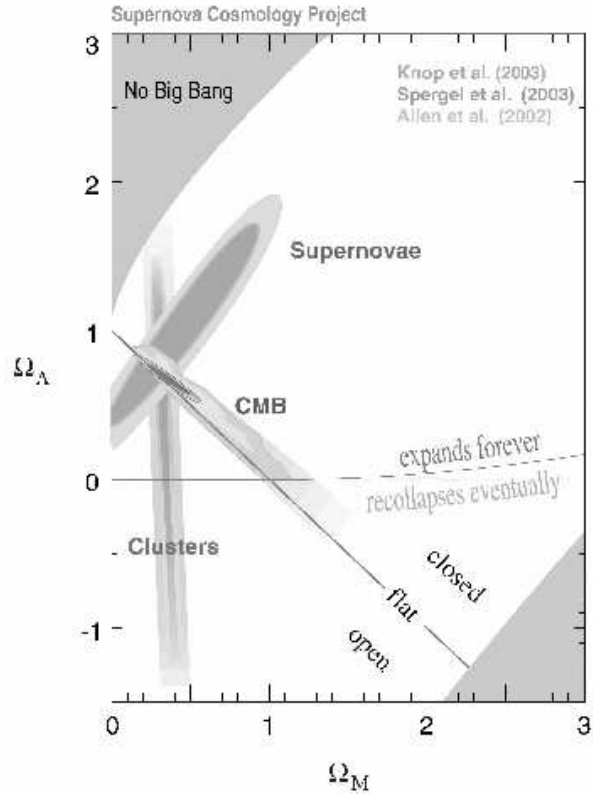}
\caption{\label{fig:lambda_matter} Confidence contours for the plane
($\Omega_m, \Omega_\Lambda$) using SN Ia, CMB and cluster density data
(taken from the Supernova Cosmology Project).}
\end{center}
\end{figure*}

The present richness of data is not exclusive of the CMB. Also large
galaxy surveys covering a large fraction of the sky and measuring the
three-dimensional power spectrum using $\approx 200000$ redshifts, like
the 2dFGRS (Percival et al. 2001) or SDSS (Tegmark et al. 2004a), have
been recently achieved. If the initial fluctuations are Gaussian, all
the information is included in the power spectrum $P(k)$ and thus this
is the quantity normally used for the analysis. The cosmological
parameters determine the matter power spectrum at present through the
initial amplitude and the spectral index $A_s, n_s$ and the transfer
function connecting linearly the present and initial $P(k)$. An
additional bias parameter $b$, assumed to be scale-independent and
linear, is needed to relate galaxy and matter density. Moreover,
non-linear evolution of the matter density power spectrum should be
also considered in the comparison with observations. The present
matter density power spectrum as well as the luminosity
distance-redshift diagram determined with SN Ia depend on the
cosmological parameters in a very different way from the $C_\ell$, and
so the combined data sets can potentially impose much more severe
constraints. Results of combining CMB data with LSS galaxy surveys
(Percival et al. 2001, Tegmark et al. 2004a), HST key project value of
$H_0$ (Freedman et al. 2001), SN Ia magnitude-redshift data (Riess et
al. 2001, Tonry et al. 2003), Ly$\alpha$ forest power spectrum (Croft
et al. 2002, McDonald et al. 2004) are given in Bennett et al. (2003),
Spergel et al. 2003, Tegmark et al. (2004b) and Seljak et
al. (2004). As it is shown in these papers, the constraints on the
cosmological parameters are greatly improved when combining all those
cosmological data sets and, what is even more important, the best fit
model is an acceptable fit for all of them. Table \ref{table:WMAP_LSS} shows
the best fit cosmological parameters given in Bennett et al. (2003)
for the combination WMAP+CBI+ACBAR+2dFGRS. Alternative combinations
include WMAP, high redshift SN Ia and abundances of rich clusters of
galaxies (see e.g. Rapetti, Steven and Weller 2005 and Jassal, Bagla
and Padmanabhan 2005 for a detailed analysis of the constraints
imposed by these data sets on the dark energy equation of state). In
figure \ref{fig:lambda_matter} confidence contours are given for the
($\Omega_m, \Omega_{DE}$) plane combining WMAP, SN Ia (Knop et
al. 2003) and galaxy cluster abundance (Allen et al. 2002). The
complementarity of the three data sets is clearly noticed by the
reduction of the contours of the combined data set compared to the
individual ones.

\begin{table*}[ht]
\begin{center}
\caption[]{\label{table:WMAP_LSS} Cosmological parameters from WMAP,
CBI, ACBAR and 2dFGRS combined data (from Bennett et al. 2003)}
\begin{tabular}{cc}
\sphline
\it Parameter& \it Values (68\% CL)\cr
\sphline
$w_b$ & 0.0224$\pm$0.0009\cr
$w_m$ & 0.135$_{-0.009}^{+0.008}$\cr
$w_\nu$ & <0.0076 (95\% CL)\cr
$w$ & <-0.78 (95\% CL)\cr
$\Omega_{DE}$ & 0.73$\pm$0.04\cr
$h$ & 0.71$_{-0.03}^{+0.04}$\cr
$\tau$ & 0.17$\pm$0.04\cr
$A_s$ & 0.833$_{-0.083}^{+0.086}$\cr
$n_s$ & 0.93$\pm$0.03\cr
$\alpha$ & -0.031$_{-0.018}^{+0.016}$\cr
$r$ & <0.90 (95\% CL)\cr
\sphline
\end{tabular}
\end{center}
\end{table*}

\subsection{Integrated Sachs-Wolfe effect}

There are strong evidences that the universe today is dominated by the
dark energy density $\Omega_{DE}$. The evidences came first from
measurements of the luminosity curve and redshift of distance SN Ia,
and later from surveys of the CMB anisotropy and LSS distribution of
galaxies. More recently there have been independent tests that confirm
this result. They come from the cross-correlation of the CMB map with
LSS surveys which span a wide range in redshift. From the late ISW
effect a non null signal is expected due to the fact that the same
gravitational potential created by the LSS at late times is also
leaving an imprint on the CMB anisotropies. The amplitude and sign of
the effect are determined by $\Omega_{DE}$, $\Omega_k$ and $h$. If the
universe is close to spatially flat, as found in a consistent way by
very different cosmological data sets, then a positive LSS-CMB
cross-correlation would be an unambiguous indication of the presence of
dark energy.

\begin{figure*}[h]
\begin{center}
\includegraphics[width=8cm]{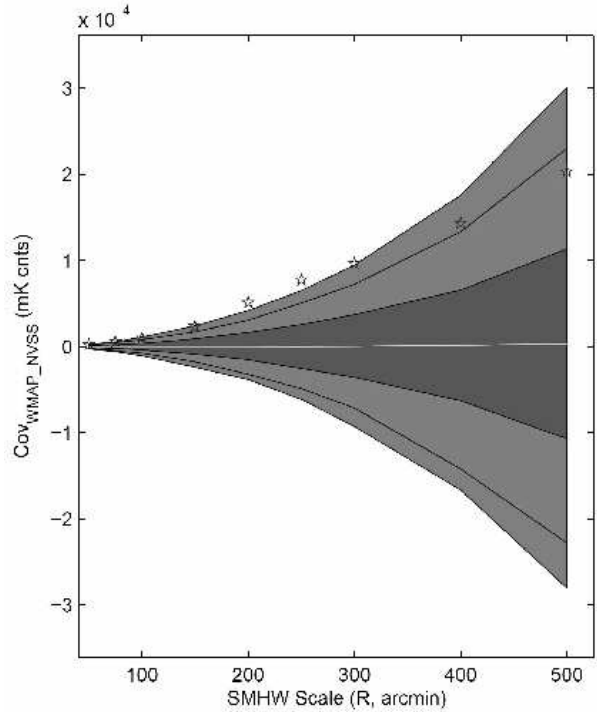}
\caption{\label{fig:ISW}Detection of the ISW effect, above $3\sigma$ level, 
applying spherical wavelets to the NVSS and WMAP maps and computing
the covariance of the wavelet coefficients as a function of the
wavelet scale. The contours are 1, 2 and 3 $\sigma$ 
(from Vielva, Martínez-González and Tucci 2004).}
\end{center}
\end{figure*}

This is what has been recently found by a number of authors using
different surveys and methods (Boughn and Crittenden 2004; Fosalba,
Gaztañaga and Castander 2004; Nolta et al. 2004; Afshordi, Loh and
Strauss 2004; Vielva, Martínez-González and Tucci 2004). The maximum
amplitude for the signal has been obtained in the latter paper by
cross-correlating the radio galaxy survey NRAO VLA Sky Survey (NVSS,
Condon et al. 1998) with the WMAP data. It represents a detection $>3\sigma$
(see figure  \ref{fig:ISW}). 

\section{Open problems} 
\label{sec:open_problems}

Looking at table \ref{table:WMAP_LSS} and considering the situation
only ten years ago, the progress in the determination of the
cosmological parameters has been spectacular. However, what is even
more impressive is the concordance among all cosmological data sets
available at present. This agreement in the results has made possible
to build a well specified model of the universe, the so called
concordance model. Nevertheless there are still pending problems
related to the data fitting as well as the interpretation of the
values of some of the parameters, whose solutions will need to wait
for future more sensitive experiments (at least for some of them, see
below).

In spite of the very precise overall fit of the theoretical $C_\ell$
to the WMAP data (Spergel et el. 2003), there are some significant
discrepancies. There are also significant deviations from the assumed
isotropic Gaussian random field hypothesis, property which is also assumed in
the parameter estimation. Moreover, some parameters take uncomfortably
high values and the nature of others is completely unknown. More
specifically, some of the most relevant open problems are:

\begin{itemize}

\item {\sl Anomalies at the lowest multipoles ($\ell=2,3$)}: deficit
of power and the coincidence of orientations of the quadrupole and
octopole in the WMAP data (Bennett et al. 2003, de Oliveira-Costa et
al. 2004, Efstathiou et al. 2004). A better removal of the Galactic
contamination narrowing the mask used for the analysis (as expected
from the future Planck mission, see below) can help to clarify this
situation.  An attempt to solve this problem invokes a global
non-trivial topology for the universe (Luminet et al. 2003).

\item {\sl Anomalies at intermediate multipoles ($\ell = 22, 40,
210$)}: glitches have
been found around those multipoles in the WMAP data (Bennett et al
2003). It is not clear if they are just statistical fluctuations,
foreground contamination or represent real features. Some of them
might be related to the asymmetries discussed below (Hansen et al. 2004).

\item {\sl Asymmetries and non-Gaussianity}: a number of authors have
found significant deviations from isotropy and Gaussianity in the WMAP
data. In particular asymmetries along the sky and/or non-Gaussianity
have been found in the genus (Park 2004), low-order moments (Eriksen
et al. 2004, Hansen et al. 2004), wavelets (Vielva et al. 2004,
Mukherjee and Wang 2004), phases (Chiang et al. 2003), extrema (Larson
and Wandelt 2004), directional wavelets (McEwen et al. 2005) and local
curvature (Cabella et al. 2005). In addition a very cold spot has been
detected and localized (Cruz et al. 2005). Although some of these
results could be due to foregrounds others seem to be very difficult
to explain in that way. Anisotropic models for the universe of the
type Bianchi VII$_{\sl h}$ with very small vorticity and shear seem to
provide a good fit to all those features (Jaffe et al. 2005); however
they require a total $\Omega=0.5$, which deviates significantly from
the value of the concordance model.

\item {\sl Power excess at high multipoles ($\ell > 2000$)}: several
high resolution experiments have found excess of power at multipoles
$2000<\ell<3000$ (CBI, DASI, ACBAR). This excess can be associated, at
least partially, to SZ emission from galaxy clusters (Bond et
al. 2005) or to extragalactic sources (Toffolatti et al 2005).  Diego
et al. (2004) suggested a sensitive way to separate between those
possibilities using the skewness and kurtosis of the Mexican Hat
wavelet coefficients.

\item {\sl High value of the optical depth}: from the WMAP data a high
value of the Thompson optical depth is derived ($\tau \simeq 0.17$,
see tables \ref{table:WMAP} and \ref{table:WMAP_LSS}), implying a high
redshift for the reionization of the inter-galactic medium of the
universe $z\sim 20$ and, thus, that non-linear structures were already
formed at these high redshifts.  This result heavily relies on the
high value of the TE angular cross-power spectrum at low multipoles
measured by WMAP, and so a confirmation with the second and other
years of data ---particularly with the E-mode angular power spectrum---
would be crucial. Again, sensitive polarization data covering a large
fraction of the sky (as expected from Planck) would represent an
important step forward to solve this problem.

\item {\sl Non-zero running index $\alpha$}: WMAP data combined
with other cosmological data sets slightly favours a running index
different from zero (see table \ref{table:WMAP_LSS}. This result is
heavily supported by the data at small scales provided by the Ly$\alpha$
forest power spectrum (Croft et al. 2002). However, new analyses of
the matter power spectrum from the Ly$\alpha$ forest find values of
$\alpha$ perfectly consistent with zero (Seljak et al. 2005).

\item {\sl Isocurvature fluctuations}: Although the WMAP temperature and
polarization data imply that the primordial density fluctuations were
primarily adiabatic (also confirmed by the recent polarization
measurements from CBI), however at the 2$\sigma$ CL up to $\approx
50\%$ of the $C_{\ell}$ power can still be due to isocurvature
fluctuations (Crotty, Lesbourgues and Pastor 2003). In addition,
considering all possible combinations of adiabatic and isocurvature
modes for each matter component will significantly lower the
constraints in the parameters (Bucher, Moodley and Turok 2001). Future
sensitive experiments like Planck will be able to further constraint
the type of the fluctuations and their relative amounts.

\item {\sl Topological defects}: the acoustic oscillations measured in
the $C_{\ell}$ by WMAP and other CMB experiments strongly reduce the
possible role of topological defects in the universe. Recently, Bevis,
Hindmarsh and Kunz (2004) and Fraisse (2005) have limited 
the contribution of topological defects (textures and cosmic strings)
to the $C_{\ell}$ to be $\lesssim 20\%$. However, a fine
determination of their possible sub-dominant contribution will probably
have to wait for future more precise data.

\item {\sl Nature of dark matter and dark energy}: probably the most
challenging problem in cosmology is to understand the nature of the
dark energy and the dark matter which completely dominates the
dynamics of the universe. For the former, a cosmological constant
(with equation of state $w=-1$) seems to be in acceptable agreement
with many different data sets. This situation is particularly
embarrassing since a natural explanation in terms of the vacuum energy,
given by quantum zero-point fluctuations of fundamental fields,
implies a value around 100 orders of magnitude above the observed
one. Regarding the dark matter, although several candidates have been
proposed within particle physics theories, neither its nature
nor its evolutionary history are known.

\item {\sl B-mode polarization}: detection of this mode would
unambiguously indicate the existence of a primordial background 
of gravitational waves and thus would allow to constrain theories of the early
universe. Due to its expected weak signal respect to the temperature,
the E-mode anisotropies (see figure \ref{fig:spectra_model}) and to the
emission from the Galactic and extragalactic foregrounds (Tucci et
al. 2005), its detection and precise determination will necessarily
require the development of new very sensitive experiments.

\end{itemize}

On top of these problems there are others, more fundamental ones,
related to the origin of inflation and to its dynamics: which is the
particle physics framework for inflation?, how the universe started to
inflate in the early universe?, which is the
specific model of inflation within, e.g., the simple classification by
Kinney (2002)?  The ESA Planck mission, the third space mission after
COBE and WMAP, scheduled for launch in 2007, will measure the
temperature of the CMB over the whole sky with high sensitivity (few
$\mu$K), resolution (down to 5 arcmin) and wide frequency coverage
($30-850$ GHz), allowing a good separation between CMB and foregrounds
and thus providing an angular power spectrum with unprecedented
precision limited only by cosmic variance. It will also provide a map
of polarization with sensitivity beyond the one reached by previous
experiments. These new measurements will certainly have a profound
impact on our understanding of the origin and evolution of our
universe (for more details on the Planck mission see the web page
address given in the references). An open issue which may remain even
after Planck is the detection of the B-mode polarisation. As stated
above, the determination of this mode would have a tremendous impact
on the theories of the early universe. There are now many plans to
build experiments capable to reach the demanding sensitivities needed
to search for the B-mode including space missions from ESA and NASA.

Finally, from the previous discussion we can state that the role
played by the CMB on our understanding of the universe has been very
relevant in the past and is expected to continue being so in the future.

\begin{acknowledgments}
We thank R.B. Barreiro for useful comments on the manuscript.  We
acknowledge financial support from the Spanish MEC project
ESP2004-07067-C03-01.  We also acknowledge the use of LAMBDA, support
for which is provided by by the NASA Office of Space Science. We
acknowledge the use of the software package CMBFAST
(http://www.cmbfast.org) developed by Seljak and Zaldarriaga. The work
has also used the software package HEALPix
(http://www.eso.org/science/healpix) developed by K.M. Gòrski,
E.F. Hivon, B.D. Wandelt, J. Banday, F.K. Hansen and M. Barthelmann.
\end{acknowledgments}


\begin{chapthebibliography}{1}

\bibitem[]{Afshordi} {Afshordi N., Loh Y.-S. \& Strauss M. A., 2004, Phys. Rev. D, 69, 3524}
%
\bibitem[]{}{Allen A.W., Schmidt R.W. \& Fabian A.C., 2002, MNRAS, 334, L11}
%
\bibitem[]{}{Barreiro R.B., 2000, New Astronomy Reviews, 44, 179}
%
\bibitem[]{}{Bartolo N., Komatsu E., Matarrese S. \& Riotto A., 2004, Phys. Rep., 402, 103}
%
\bibitem[]{Bennett} {Bennett C. L. et al., 2003, ApJS, 148, 1}
%
\bibitem[]{}{Bersanelli M., Maino, D. \& Mennella A., 2002, Nuovo Cimento, 25, 1}
%
\bibitem[]{}{Bevis N., Hindmarsh M. \& Kunz M., 2004,
Phys. Rev. D, 70, 043508}
%
\bibitem[]{Bond} {Bond et al., 2005, ApJ, 626, 12}
%
\bibitem[]{}{Bouchet F.R. \& Gispert R., 1999, New Astronomy Reviews, 4, 443}
%
\bibitem[]{Boughn} {Boughn S. P. \& Crittenden R. G., 2004, Nature, 427, 45}
%
\bibitem[]{}{Bucher M., Moodley K. \& Turok N., 2001,
Phys. Rev. Lett., 87, 191301}
%
\bibitem[]{}{Cabella, P. \& Kamionkowski, M., 2004, astro-ph/0403392}
%
\bibitem[]{}{Cabella P., Liguori M., Hansen F.K., Marinucci D., Matarrese S., 
Moscardini L. \& Vittorio N., 2005, MNRAS, 358, 648}
%
\bibitem[]{Cruz} {Cruz M., Martínez-González E., Vielva P. \& Cayón L., 2005, 
MNRAS, 356, 29}
%
\bibitem[]{}{Challinor A., 2004, astro-ph/0403344}
%
\bibitem[]{Challinor} {Challinor A., 2005, astro-ph/0502093}
%
\bibitem[]{}{Chiang L.-Y., Naselsky P.D., Verkhodanov O.V. \& Way
M.J., 2003, ApJL, 590, 65}
%
\bibitem[]{}{Condon J.J. et al., 1998, AJ, 115, 1693}
%
\bibitem[]{}{Croft R.A.C. et al., 2002, ApJ, 581, 20}
%
\bibitem[]{}{Crotty P., Lesbourgues J. \& Pastor S., 2003,
Phys. Rev. D, 67, 123005}
%
\bibitem[]{de Bernardis} {De Bernardis P. et al., 2000, Nature, 404,
955}
%
\bibitem[]{}{de Oliveira-Costa A., Tegmark M., Zaldarriaga M. \& Hamilton A.,
2004, Phys. Rev. D, 69, 063516}
%
\bibitem[]{}{Delabrouille J., Cardoso J.-F. \& Patanchon G., 2003, MNRAS, 
346, 1089}
%
\bibitem[]{}{Dickinson C. et al., 2004, MNRAS, 353, 732}
%
\bibitem[]{}{Diego J.M., Martínez-González E., Vielva P. \& Silk J., 2004, 
astro-ph/0403561}
%
\bibitem[]{}{Efstathiou G., 2004, MNRAS, 348, 885}
%
\bibitem[]{}{Eriksen H.K., Hansen F.K., Banday A.J., Gorski K.M. \&
Lilje P.B., 2004, ApJ, 605, 14}
%
\bibitem[]{Fosalba} {Fosalba P., Gaztañaga E. \& Castander F., 2004, ApJL, 
597, 89} 
%
\bibitem[]{}{Fraisse A.A., 2005, Phys. Rev. Lett., submitted, astro-ph/0505402}
%
\bibitem[]{}{Freedman W.L. et al., 2001, ApJ, 553, 47}
%
\bibitem[]{Hanany}{Hanany S. et al., 2000, ApJ, 545, L5}
%
\bibitem[]{}{Hansen F.K., Banday A.J. \& Gòrski K.M., 2004, MNRAS,
354, 641}
%
\bibitem[]{}{Herranz D., Sanz J.L., Hobson M.P., Barreiro R.B., Diego J.M., 
Martínez-González E. \& Lasenby A.N., 2002, MNRAS, 336, 1057}
%
\bibitem[]{}{Hinshaw G. et al., 2003, ApJ, 148, 135}
%
\bibitem[]{}{Hu W. and Dodelson S., 2002, ARAA, 40, 171}
%
\bibitem[]{}{Hobson M.P., Jones A.W., Lasenby A.N. \& Bouchet F., 1998, 
MNRAS, 300}
%
\bibitem[]{}{Jaffe T.R., Banday A.J., Eriksen H.K., Gorski K.M. \& Hansen 
F.K.,2005, ApJL, submitted, astro-ph/0503213}
%
\bibitem[]{}{Jassal H.K., Bagla J.S. \& Padmanabhan T., 2005, astro-ph/0506748}
%
\bibitem[]{}{Kamionkowski M., Kosowsky A. \& Sttebins A., 1997, Phys. Rev. D, 
55, 7368}
%
\bibitem[]{}{Kinney W.H., 2002, Phys. Rev. D, 66, 083508}
%
\bibitem[]{}{Knop R.A. et al., 2003, ApJ, 598, 102}
%
\bibitem[]{}{Kogut A. et al., 2003, ApJS, 148, 161}
%
\bibitem[]{}{Kovac J.M. et al. 2002, Nature, 420, 772}
%
\bibitem[]{Komatsu} {Komatsu E. et al., 2004, ApJS, 148, 119}
%
\bibitem[]{}{Kuo C.L. et al., 2004, ApJ, 600, 32}
%
\bibitem[]{}{Larson D.L. \& Wandelt B.D., 2004, ApJ, 613, 85}
%
\bibitem[]{}{Leitch et al., 2004, ApJ, submitted, astro-ph/0409357}
%
\bibitem[]{}{Luminet J.-P., Weeks J.R., Riazuelo A., Lehoucq R. \&
Uzan J.-P., 2003, Nature, 425, 593}
%
\bibitem[]{}{Maino D. et al., 2002, MNRAS, 334,53}
%
\bibitem[]{}{Martínez-González E., Sanz J.L. \& Silk, 1990, ApJ, 335, 5}
%
\bibitem[]{}{Martínez-González E., Diego J.M., Vielva P. \& Silk J., 2003, 
MNRAS, 345, 1101}
%
\bibitem[]{}{Mather J.C. et al. 1994, ApJ, 420, 439}
%
\bibitem[]{}{Mather J.C. et al. 1999, ApJ, 512, 511}
%
\bibitem[]{}{McDonald P. et al., 2004, ApJ, submitted (astro-ph/0407377)}
%
\bibitem[]{McEwen} {McEwen J. D., Hobson M. P., Lasenby A. N. \& Mortlock D. 
J., 2005, MNRAS, 359, 1583}
%
\bibitem[]{}{Mennella A. et al., 2004, astro-ph/0402528}
%
\bibitem[]{Mukherjee} {Mukherjee P. \& Wang Y., 2004, ApJ, 613, 51}
%
\bibitem[]{Nolta} {Nolta M. R. et al., 2004, ApJ, 608, 10}
%
\bibitem[]{}{Park C.-G., 2004, MNRAS, 349, 313}
%
\bibitem[]{}{Percival W.J. et al., 2001, MNRAS, 327, 1297}
%
\bibitem[]{}{Planck mission, http://www.rssd.esa.int/Planck}
%
\bibitem[]{}{Rapetti D., Steven S.W. \& Weller J., 2005, MNRAS, 360, 555}
%
\bibitem[]{}{Readhead A.C.S. et al., 2004a, ApJ, 609, 498}
%
\bibitem[]{Readhead} {Readhead A.C.S. et al., 2004b, Science, 306, 836}
%
\bibitem[]{}{Riess et al., 2001, ApJ, 560, 49}
%
\bibitem[]{ISW} {Sachs R. K. \& Wolfe A. M., 1967, ApJ, 147, 73}
%
\bibitem[]{}{Sanz J.L., 1997, In The Cosmic Microwave Background, eds. C.H. 
Lineweaver et al., Kluwer Academic Publishers, p. 33}
%
\bibitem[]{}{Scott D.H., Srednicki M. \& White M., 1994, ApJ, 421, 5}
%
\bibitem[]{CMBFAST} {Seljak U. \& Zaldarriaga M., 1996, ApJ, 469, 437}
%
\bibitem[]{}{Seljak U. et al., 2005, Phys. Rev. D, 71, 103515}
%
\bibitem[]{}{Silk J., 1968, ApJ, 151, 459}
%
\bibitem[]{DMR} {Smoot G. F. et al., 1992, ApJL, 396, L1}
%
\bibitem[]{WMAP:parameters} {Spergel D. N. et al., 2003, ApJS, 148, 175}
%
\bibitem[]{}{Sunyaev R.A. \& Zeldovich Y.B., 1972, Comm. Astrophys. Space 
Phys., 4, 173}
%
\bibitem[]{}{Tegmark M., Eisenstein D.J., Hu W. \& de Oliveira-Costa A., 
2000, ApJ, 530, 133}
%
\bibitem[]{}{Tegmark M. et al., 2004a, ApJ, 606, 702}
%
\bibitem[]{}{Tegmark M. et al., 2004b, Phys. Rev. D, 69, 103501}
%
\bibitem[]{}{Tonry J.L. et al., 2003, ApJ, 594, 1}
%
\bibitem[]{}{Tucci M., Martínez-González E., Vielva P. \& Delabrouille J., 
2005, MNRAS, 360, 935}
%
\bibitem[]{}{Toffolatti L., Negrello M., González-Nuevo J., de Zotti
G., Silva L., Granato G.L. \& Argüeso F., 2004, A\&A, in press, 
astro-ph/0410605}
%
\bibitem[]{}{Vielva P., Barreiro R. B., Hobson M. P., Martínez-González E., 
Lasenby A. N., Sanz J. L. \& Toffolatti L., 2001, MNRAS, 328, 1}
%
\bibitem[]{Vielva1} {Vielva P., Martínez-González E., Barreiro R.B., Sanz 
J.L. \& Cayón L., 2004, ApJ, 609, 22}
%
\bibitem[]{Vielva2} {Vielva P., Martínez-González E. \& Tucci M., 2004,
MNRAS, submitted, astro-ph/0408252}
%
\bibitem[]{}{Watson R.A., Rebolo R., Rubiño-Martín J.A., Hildebrant
S., Gutiérrez C.M., Fernádez-Cerezo S., Hoyland R.J. \& Battistelli
E.S., 2005, ApJ, submitted, astro-ph/0503714}
%
\bibitem[]{}{Zaldarriaga M. \& Seljak U., 1997, Phys. Rev. D, 55, 1830}
%

\end{chapthebibliography}
\end{document}